\RequirePackage{fix-cm}
\documentclass[twocolumn]{svjour3}          
\smartqed  
\usepackage{graphicx}
\usepackage{subfig}
\usepackage{amsmath}
\usepackage{glossaries}
\usepackage{siunitx}
\usepackage{pgf,pgffor}
\usepackage[colorlinks=false, hidelinks=true]{hyperref}
\usepackage[misc]{ifsym}

\newacronym{NN}{NN}{neural network}
\newacronym{ROC}{ROC}{receiver operating characteristic}
\newacronym{AUC}{AUC}{area under the curve}

\begin{document}

\title{Identifying the relevant dependencies of the neural network response on characteristics of the input space}


\author{Stefan Wunsch \and Raphael Friese \and Roger Wolf \and G\"unter Quast}


\institute{
Stefan Wunsch\textsuperscript{1,2} \Letter \at
stefan.wunsch@cern.ch
\and
Raphael Friese\textsuperscript{1} \at
raphael.marius.friese@cern.ch
\and
Roger Wolf\textsuperscript{1} \at
roger.wolf@cern.ch
\and
G\"unter Quast\textsuperscript{1} \at
guenter.quast@kit.edu
\and
\textsuperscript{1} Karlsruhe Institute of Technology, Institute of Experimental Particle Physics, Karlsruhe, Germany\\
\textsuperscript{2} CERN, Geneva, Switzerland
}


\maketitle

\begin{abstract}

The relation between the input and output spaces of \glspl{NN} is investigated to identify those characteristics of the input space that have a large influence on the output for a given task.
For this purpose, the \gls{NN} function is decomposed into a Taylor expansion in each element of the input space.
The Taylor coefficients contain information about the sensitivity of the \gls{NN} response to the inputs.
A metric is introduced that allows for the identification of the characteristics that mostly determine the performance of the \gls{NN} in solving a given task.
Finally, the capability of this metric to analyze the performance of the \gls{NN} is evaluated based on a task common to data analyses in high-energy particle physics experiments.

\end{abstract}

\glsresetall

\section{Introduction}
\label{sec:introduction}

A \gls{NN} is a multi-parameter system, which, depending on its architecture, can consist of several thousands of weight and bias parameters, subject to one or more non-linear activation functions.
Each of these adjustable parameters obtains its concrete value and meaning by minimization during the training process.
Thus the same \gls{NN} can be applied to several concrete tasks, which are only defined at the training step.

In applications in high-energy particle physics, which are supposed to distinguish a signal from one or more backgrounds, the training sample is obtained either from simulation or from an independent dataset without overlap with the sample of interest, to which the \gls{NN} is applied.
Usually the \gls{NN} output itself is then subject to a detailed likelihood based hypothesis test, to infer the presence and yield of the signal~\cite{junk1999confidence,read2002presentation,atlas2011procedure,belforte2012combined,cowan2011asymptotic}.
The likelihood may include information on the shape of a variable that is supposed to discriminate signal from background.
This shape could (while it does not have to) be e.g. the output of an \gls{NN}.
Apart from one or more parameters of interest the hypothesis test may comprise several hundreds of nuisance parameters, steering the response of the test statistic on a corresponding set of uncertainties.
The nuisance parameters can be correlated or uncorrelated with the shape of the discriminating variable and (directly or indirectly) depend on the response of the \gls{NN} output on its input variables.

These kinds of analyses connect the observation of a measurement to a hypothesized truth. For \gls{NN} applications they pose the intrinsic problem that, beyond statistical fluctuations, congruency between the training sample and the sample of interest may not be given.
Deviations need to be identified and quantified within the uncertainty model of the hypothesis test.
They may occur not only in the description of single input variables to the \gls{NN}, but also in correlations across input variables, even if the marginal distributions of the individual input variables are reproduced.
An \gls{NN} can be sensitive to correlations across input variables; in fact this sensitivity is the main reason for potential performance gains, with respect to other approaches, like e.g. profile likelihoods.
To make sure that this performance gain is not feigned, in addition to the marginal distributions, all correlations across input variables need to be carefully checked, and their influence on the test statistic identified and eventually mapped into the uncertainty model of the hypothesis test.
The complexity of this methodology motivates the interest, not only in keeping the number of inputs to the \gls{NN} at a manageable level, but above all in identifying those characteristics of the input space to the \gls{NN} with the largest influence on the \gls{NN} output.
The definition of the uncertainty model of the hypothesis test can then be concentrated on these most influential characteristics.

This approach sets the scope of this study to not more than a few tenth, up to a few hundred, partially highly correlated input variables in the context of particle physics experiments, or comparable applications.
It differs from the approaches of weak supervision~\cite{metodiev2017weakly,dery2017weakly,komiske2018weakly,cohen2018weakly} and pivoting with adversaries~\cite{louppe2017learning} that have been discussed in the literature.
Weak supervision tries to circumvent the problem that we are describing by replacing an originally ground-truth labeled training by a training based on unlabeled training data.
The corresponding samples can be obtained from the data themselves.
They do not depend on a simulation and may be chosen to be unbiased.
This approach is well justified in classification tasks, that are based just on the characteristics of the predefined training data.
In the analyses that we are discussing the classification is tied to the hypothesized truth.
Replacing the ground-truth labeled training by unlabeled input data does not solve the problem that we are discussing.
Our discussion is also beyond the scope of pivoting with adversaries, for which the mismodellings to address have to be known beforehand.
Our discussion sets in at an earlier stage, which is the most complete identification of all uncertainties that can be of relevance for the physics analysis.
After the most influential features of the input space have been identified the method of pivoting with adversaries could be applied to mitigate potential mismodellings.
A related approach to extract information about the characteristics of the input space is to flatten the distributions of sub-spaces so that possible discriminating features vanish~\cite{deoliveira2017flattening,chang2018flattening}.
From the performance degradation after retraining the \gls{NN} on the modified inputs, information about the discriminating power of the respective sub-space can be obtained.
However, this approach does not allow to evaluate the dependencies of the response of an unique \gls{NN} function on the characteristics of the input space, since each retrained function may have learned different features.

So far, the questions we are raising have been addressed by methods that have been proposed to relate the output of \glspl{NN} with certain regions of input pixels in the context of image classification~\cite{bach2015pixel,montavon2017explaining}.
These methods only use first-order derivatives to the \gls{NN} function to back propagate the output layer by layer.
What we propose is a Taylor expansion of the full \gls{NN} function up to an arbitrary order, which allows to connect the input space directly to the \gls{NN} output.
While with this study we will demonstrate the application of the Taylor expansion only up to second order, we explicitly propose a generalization towards higher-order derivatives in the Taylor expansion to capture relations across variables, which usually play a more important role in data analyses in high-energy particle physics experiments.

Due to the high-performance computation of derivatives in modern software frameworks used for the implementation of \glspl{NN}~\cite{abadi2016tensorflow,paszke2017automatic,bergstra2010theano}, this expansion can be obtained at each point of the input space, even if this space is of high dimension.
In this way, the sensitivity of the \gls{NN} response to the input space can be analyzed by the gradient of the \gls{NN} function.
For practical reasons we stop the expansion at second order. To facilitate the following interpretation, we define a feature to be a characteristic of a single element or a pair-wise relation between two elements of the input space.
The first class of features relates to the coefficients of the expansion to first order (first-order feature); the second class to the coefficients of the second order expansion (second-order feature).
First-order features capture the influence of single input elements on the \gls{NN} output throughout the input space; second-order features the influence of pair-wise or auto-correlations among the input elements.
It is obvious that depending on the given task a certain feature can have large influence on the output of the \gls{NN} in a certain region of the input space, while it is less important in others.
We propose the arithmetic mean of the absolute value of the corresponding Taylor coefficient, computed from the input space defined by the task to be solved,
\begin{equation}\label{eq:metric}
\langle t_{i} \rangle \equiv \frac{1}{N} \sum_{k=1}^{N}\left|t_{i}(\left.\{x_{j}\}\right|_{k})\right|\qquad i\in\mathcal{P}(\{x_{j}\})
\end{equation}
as a metric for the influence of a given feature of the input space on the output, where the sum runs over the whole testing sample of size $N$, $t_i$ corresponds to the coefficients of the Taylor expansion, $\{x_{j}\}|_{k}$ to the set of variables spanning the input space, evaluated for element $k$ of the testing sample, and $i$ is an element of the powerset of $\{x_j\}$.
It should be noted that the $\langle t_i \rangle$ characterize the input space (as covered by the test data) and the sensitivity of the \gls{NN} to it, after training, as a whole.

In section~\ref{sec:toy_scenarios} we illustrate this choice with the help of four simple tasks emphasizing certain single features of the input space or their combination.
In section~\ref{sec:explaining_the_learning_progress} we point out that, when evaluated at each step of the minimization during the training process, the $\langle t_{i} \rangle$ can be utilized to illustrate and monitor the training process and learning strategies adopted by the \gls{NN}.
In section~\ref{sec:application_on_a_scenario_from_particle_physics} we show the application of the $\langle t_{i} \rangle$ to a more realistic task common to data analyses in high-energy particle physics experiments.
Such tasks usually have the following attributes, which are of relevance for the following discussion:
\begin{itemize}
\item they consist of not more than several tens of important input parameters, which leads to a moderate dimensionality of the posed problem;
\item they may rely on relations between elements more than they rely on single elements of the input space;
\item they usually pose problems, where a signal and background class cannot be separated based on single or few input variables, but only from the combination of several input variables;
\item they require a good understanding of the \gls{NN} performance to turn the output into a reliable measurement.
\end{itemize}

\section{Analysis of features of the input space for simple tasks}
\label{sec:toy_scenarios}

In the following we illustrate the relation of the $\langle t_i\rangle$ to certain features of the input space.

The applied \gls{NN} corresponds to a fully connected feed-forward model with a single hidden layer consisting of 100 nodes.
As activation functions a hyperbolic tangent is chosen for the hidden layer and a sigmoid for the output layer.
A preprocessing of the inputs is performed following the $(x-\mu)/\sigma$ rule with the mean $\mu$ and the standard deviation $\sigma$ derived independently for each input variable.
The free parameters of the \gls{NN} are fitted to the training data using the cross-entropy loss and the Adam optimizer algorithm~\cite{kingma2014adam}.
The full training dataset with $\SI{e5}{}$ elements is split into two equal halves.
One half is used for the calculation of the gradients used by the optimizer.
The other half is used as independent validation dataset.
The training is stopped if the loss did not improve on the validation dataset for three times in a row (early stopping).
The independent test dataset used to calculate the $\langle t_i\rangle$ consists of $\SI{e5}{}$ elements.
We use the software packages Keras~\cite{chollet2015keras} and TensorFlow~\cite{abadi2016tensorflow} for the implementation of the \gls{NN} and the calculation of the derivatives.

\begin{figure*}
\subfloat[][]{
\begin{minipage}{0.24\textwidth}
\hspace{3mm}\includegraphics[width=0.92\linewidth]{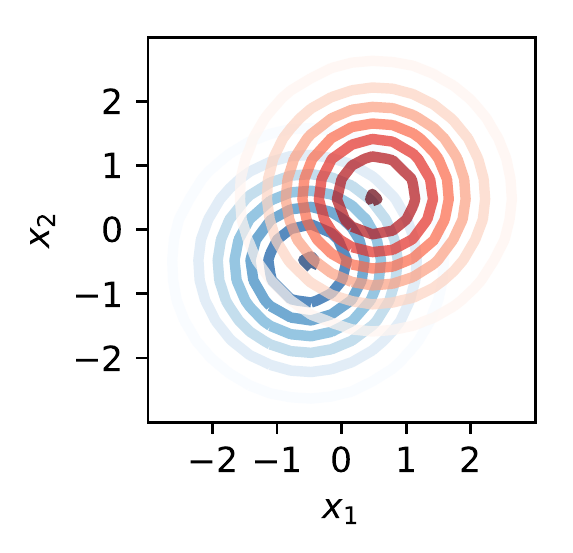}
\includegraphics[width=0.99\linewidth]{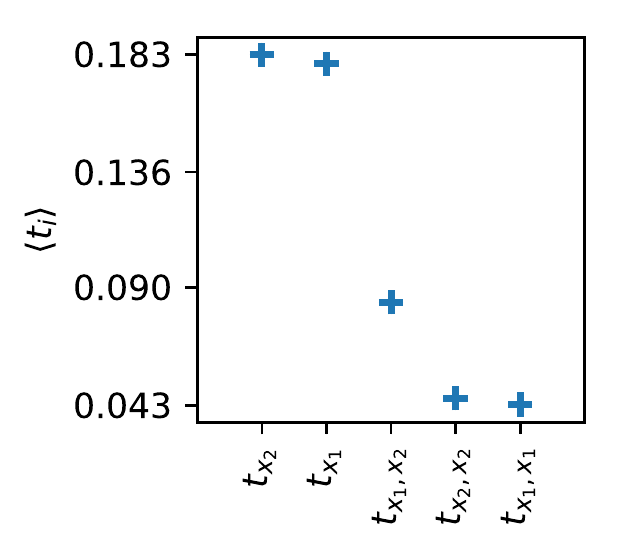}
\end{minipage}
\label{fig:scenario_a}}
\subfloat[][]{
\begin{minipage}{0.24\textwidth}
\hspace{3mm}\includegraphics[width=0.92\linewidth]{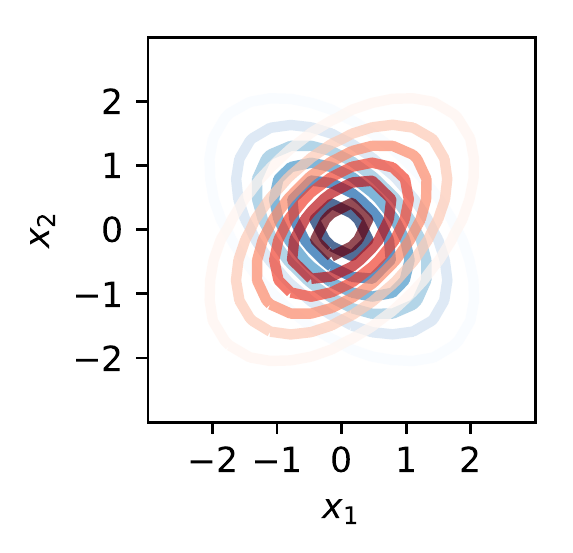}
\includegraphics[width=0.99\linewidth]{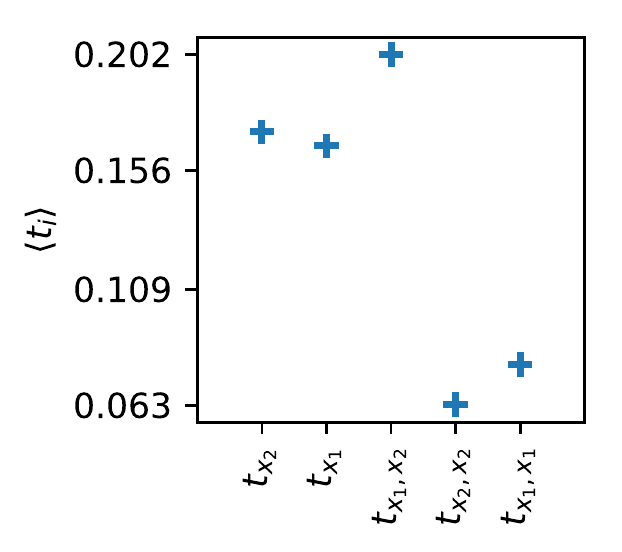}
\end{minipage}
\label{fig:scenario_b}}
\subfloat[][]{
\begin{minipage}{0.24\textwidth}
\hspace{3mm}\includegraphics[width=0.92\linewidth]{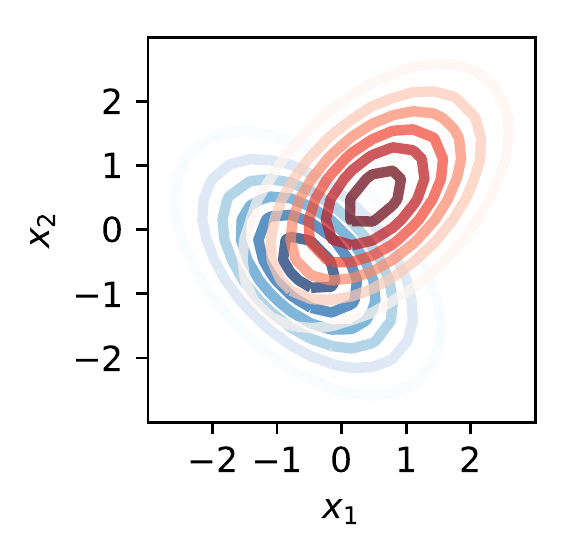}
\includegraphics[width=0.99\linewidth]{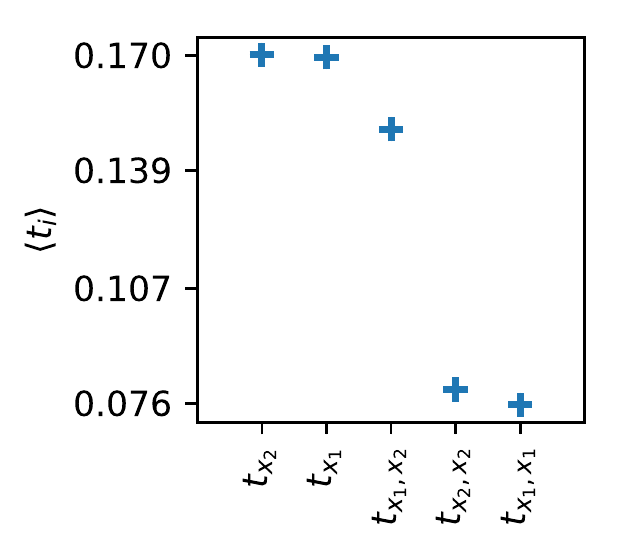}
\end{minipage}
\label{fig:scenario_c}}
\subfloat[][]{
\begin{minipage}{0.24\textwidth}
\hspace{3mm}\includegraphics[width=0.92\linewidth]{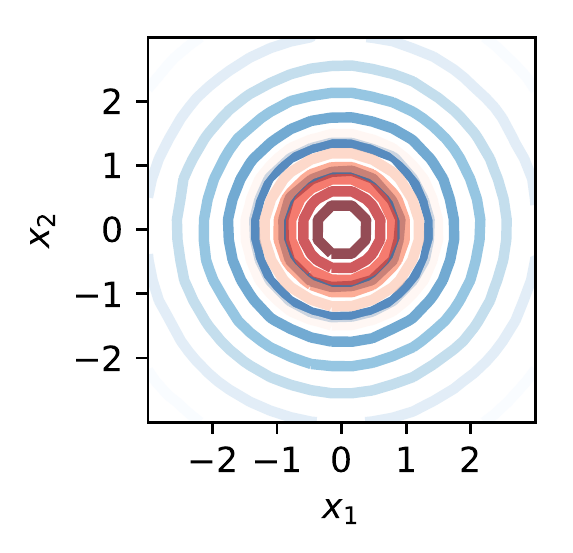}
\includegraphics[width=0.99\linewidth]{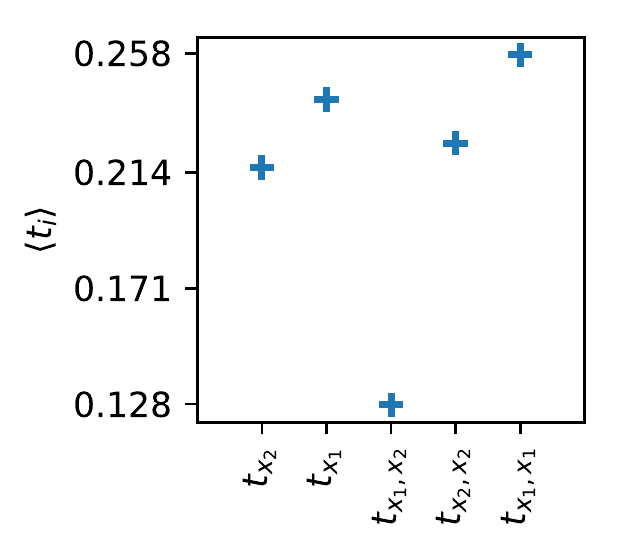}
\end{minipage}
\label{fig:scenario_d}}
\caption{(Upper row) Contours of the distributions used in the examples for the signal (red) and background (blue) classes discussed in section~\ref{sec:toy_scenarios}, and the (lower row) corresponding metrics $\langle t_i \rangle$.}
\label{fig:scenarios}
\end{figure*}

For simplicity we choose binary classification tasks with two inputs, $x_1$ and $x_2$.
For the signal and background classes we sample Gaussian distributions with parameters, as summarized in Table~\ref{tab:scenarios}.
From the Taylor series we obtain two metrics $\langle t_{x_1}\rangle$ and
$\langle t_{x_2}\rangle$ indicating the influence of the marginal distributions of $x_1$ and $x_2$, and three metrics $\langle t_{x_1,x_1}\rangle$, $\langle t_{x_1,x_2}\rangle$, and $\langle t_{x_2,x_2}\rangle$ indicating the influence of the relation between $x_1$ and $x_2$, and the two auto-correlations.
In the upper row of Fig.~\ref{fig:scenarios} the distribution of the (red) signal and (blue) background classes in the input space are shown, where darker colors indicate a higher sample density.
In the lower row of Fig.~\ref{fig:scenarios} the values obtained for the $\langle t_i \rangle$ after the training are shown for each corresponding task.

For the task shown in Fig.~\ref{fig:scenario_a} the signal and background classes are shifted against each other.
In both classes $x_1$ and $x_2$ are uncorrelated and of equal spread.
The classification task becomes most difficult along the off-diagonal axis between the two classes through the origin and simpler if both, $x_1$ and $x_2$, take large or small values at the same time.
Correspondingly, $\langle t_{x_1}\rangle$ and $\langle t_{x_2}\rangle$ obtain large values indicating the separation power that is already caused by the marginal distributions of $x_1$ and $x_2$. The orientation of the two classes with respect to each other also results in a non-negligible contribution of $\langle t_{x_1,x_2}\rangle$ to the \gls{NN} response.

For the task shown in Fig.~\ref{fig:scenario_b} the signal and background classes are both centered at the origin of the input space, with equal spread in $x_1$ and $x_2$, but with different correlation coefficients in the covariance matrix.
The classification task is most difficult in the origin of the input space and becomes simpler if $x_1$ and $x_2$ take large absolute values.
Correspondingly, the relation between $x_1$ and $x_2$ is identified as the most influential feature by the value of $\langle t_{x_1,x_2} \rangle$.
The fact that large absolute values of $x_1$ and $x_2$ support the separability of the two classes is expressed by the relatively large values for $\langle t_{x_1} \rangle$ and $\langle t_{x_2}\rangle$.
A combination of the examples of Fig.~\ref{fig:scenario_a} and \ref{fig:scenario_b} is shown in Fig.~\ref{fig:scenario_c}.
For the task shown in Fig.~\ref{fig:scenario_d} the signal and background classes are both centered in the origin of the input space with different spread.
In both classes $x_1$ and $x_2$ are uncorrelated.
According to the symmetry of the posed problem the relation between $x_1$ and $x_2$ is expected to not strongly contribute to the separability of the signal and background classes.
This is confirmed by the lower value of $\langle t_{x_1,x_2} \rangle$.
Instead $\langle t_{x_1} \rangle$, $\langle t_{x_2} \rangle$, $\langle t_{x_1,x_1} \rangle$, and $\langle t_{x_2,x_2} \rangle$ take larger values as expected from the previous discussion.

\begin{table*}
  \begin{center}
   \caption{
Parameters defining the signal and background classes used for the tasks discussed in section~\ref{sec:toy_scenarios}. The parameters correspond to two-dimensional Gau{\ss}ian distributions.}
\label{tab:scenarios}
    \begin{tabular}{lcccccc}
    Task &
    \multicolumn{4}{c}{Mean value} &
    \multicolumn{2}{c}{Covariance matrix} \\
    &
    \multicolumn{2}{c}{Signal ($x_{1}$, $x_{2}$)} &
    \multicolumn{2}{c}{Background ($x_{1}$, $x_{2}$)}  &
    Signal &
    Background  \\
    \hline
        Fig.~\ref{fig:scenario_a} &
    \hphantom{-}0.5 &
    \hphantom{-}0.5 &
    $-0.5$ &
    $-0.5$ &
    $\vphantom{\Biggl(\Biggr)}\left(\begin{array}{cc} \hphantom{-}1\hphantom{.5} & \hphantom{-}0\hphantom{.5} \\
    \hphantom{-}0\hphantom{.5} & \hphantom{-}1\hphantom{.5}\end{array}\right)$ &
    $\vphantom{\Biggl(\Biggr)}\left(\begin{array}{cc} \hphantom{-}1\hphantom{.5} & \hphantom{-}0\hphantom{.5} \\
    \hphantom{-}0\hphantom{.5} & \hphantom{-}1\hphantom{.5}\end{array}\right)$ \\
        Fig.~\ref{fig:scenario_b} &
    \hphantom{-}0\hphantom{.0} &
    \hphantom{-}0\hphantom{.0} &
    $\hphantom{-}0\hphantom{.0}$ &
    $\hphantom{-}0\hphantom{.0}$ &
    $\vphantom{\Biggl(\Biggr)}\left(\begin{array}{cc} \hphantom{-}1\hphantom{.5} & \hphantom{-}0.5 \\
    \hphantom{-}0.5 & \hphantom{-}1\hphantom{.5}\end{array}\right)$ &
    $\vphantom{\Biggl(\Biggr)}\left(\begin{array}{cc} \hphantom{-}1\hphantom{.5} & -0.5 \\
    -0.5 & \hphantom{-}1\hphantom{.5}\end{array}\right)$ \\
        Fig.~\ref{fig:scenario_c} &
    \hphantom{-}0.5 &
    \hphantom{-}0.5 &
    $-0.5$ &
    $-0.5$ &
    $\vphantom{\Biggl(\Biggr)}\left(\begin{array}{cc} \hphantom{-}1\hphantom{.5} & \hphantom{-}0.5 \\
    \hphantom{-}0.5 & \hphantom{-}1\hphantom{.5}\end{array}\right)$ &
    $\vphantom{\Biggl(\Biggr)}\left(\begin{array}{cc} \hphantom{-}1\hphantom{.5} & -0.5 \\
    -0.5 & \hphantom{-}1\hphantom{.5}\end{array}\right)$ \\
        Fig.~\ref{fig:scenario_d} &
    \hphantom{-}0\hphantom{.0} &
    \hphantom{-}0\hphantom{.0} &
    $\hphantom{-}0\hphantom{.0}$ &
    $\hphantom{-}0\hphantom{.0}$ &
    $\vphantom{\Biggl(\Biggr)}\left(\begin{array}{cc} \hphantom{-}0.5 & \hphantom{-}0\hphantom{.5} \\
    \hphantom{-}0\hphantom{.5} & \hphantom{-}0.5\end{array}\right)$ &
    $\vphantom{\Biggl(\Biggr)}\left(\begin{array}{cc} \hphantom{-}3\hphantom{.5} & \hphantom{-}0\hphantom{.5} \\
    \hphantom{-}0\hphantom{.5} & \hphantom{-}3\hphantom{.5}\end{array}\right)$ \\
    \end{tabular}
  \end{center}
\end{table*}

\section{Analysis of the learning progress}
\label{sec:explaining_the_learning_progress}

\begin{figure}
\centering
\includegraphics[width=0.92\linewidth]{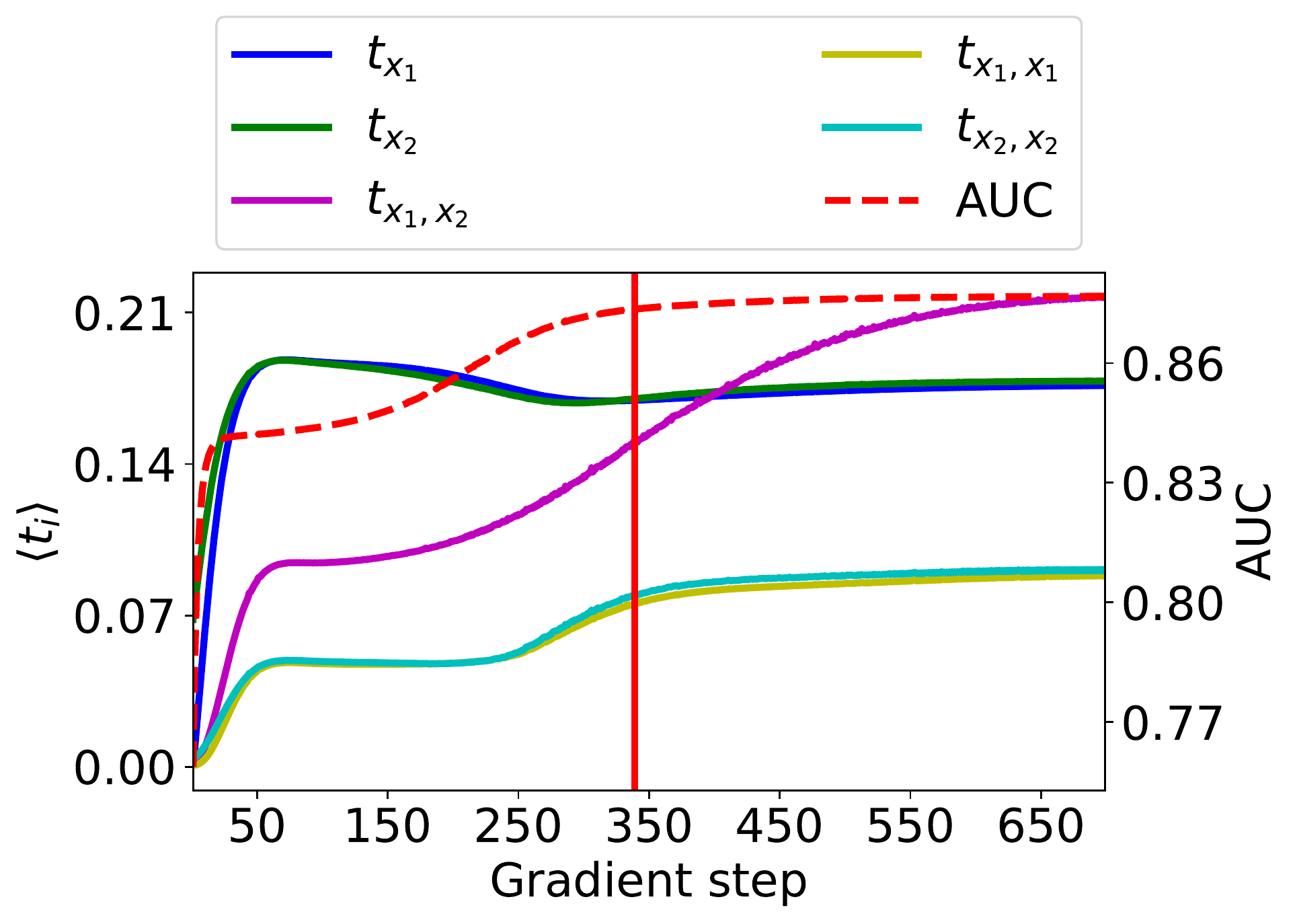}
\caption{Values of the metrics $\langle t_i \rangle$, as defined in Eq.~\ref{eq:metric}, evaluated at each gradient step of the \gls{NN} training, for the task discussed in section~\ref{sec:toy_scenarios} and shown in Fig.~\ref{fig:scenario_c}. On the axis to the right the \gls{AUC} of the \gls{ROC} curve, as a measure of the \gls{NN} performance in solving the task at each training step, is shown.
The red vertical line indicates after how many gradient steps the predefined stopping criterion, given in section~\ref{sec:toy_scenarios}, has been met.}
\label{fig:animated}
\end{figure}

When evaluated at each minimization step during the
training, the metrics $\langle t_i \rangle$ may serve as a tool to analyze the learning progress of the \gls{NN}.
We illustrate this for the task shown in Fig.~\ref{fig:scenario_c}.
In Fig.~\ref{fig:animated} the evolving values of each $\langle t_i \rangle$ are shown, as continuous lines of different color, for the first 700 gradient steps. The stopping criterion of the training is reached after 339 gradient steps (indicated by the red vertical line in the figure).
We measure the performance of the \gls{NN} in separating the signal from the background class by the \gls{AUC} of the \gls{ROC}.
We have added the \gls{AUC} at each training step to the figure with a separate axis on the right.
A rough distinction of two phases can be stated.
Approximately up to minimization step 30 the performance of the \gls{NN} shows a steep rise up to a plateau value of $0.84$ for the \gls{AUC}.
This rise coincides with increasing values of $\langle t_{x_1}\rangle$ and $\langle t_{x_2}\rangle$.
Both metrics have the same progression, which can be explained by the symmetry of the task.
Also the values for $\langle t_{x_1,x_1}\rangle$, $\langle t_{x_1,x_2}\rangle$ and $\langle t_{x_2,x_2}\rangle$ show an increase, though much less pronounced.
Roughly 100 minimization steps later, a second, more shallow, rise of the \gls{AUC} sets in, coinciding with increasing values for $\langle t_{x_1,x_2}\rangle$.
We interpret this in the following way.
During the first phase the \gls{NN} adapts to the first-order features related to $\langle t_{x_1}\rangle$ and $\langle t_{x_2}\rangle$, which is the most obvious choice to separate the signal from the background class.
During this phase the learning progress of the \gls{NN} is concentrated in the areas of the input space with medium to large values of $x_1$ and $x_2$.
In the second phase the relation between $x_1$ and $x_2$, as a second-order feature, gains influence.
This is when the \gls{NN} learning progress concentrates on the region of the input space where the signal and background classes overlap.
It can be seen that the influence of the features related to $\langle t_{x_1}\rangle$ and $\langle t_{x_2}\rangle$ decreases from minimization step 50 on.
Apparently this influence has been overestimated at first and is successively replaced giving more importance to the more difficult to identify second-order features.
From our knowledge of the truth, this is indeed the "more correct" assessment, which from
minimization step 250 on, also leads to another gain in performance.
Note that by the end of the training the progression of $\langle t_{x_1,x_2}\rangle$ has not converged, yet.
The stopping criterion represents a measure of success and not a measure of truth.
It might well have happened that the stopping criterion might have been met already between gradient step 50 and 100.
In this case the \gls{NN} output would have been based on the assessment that $\langle t_{x_1,x_2}\rangle$ plays a less important role.
In this case success rules over truth. In our example the a priori known, more correct assessment leads to another performance gain after a few more gradient steps.
Stopping the training before gradient step 100 would have missed this performance gain.
We would like to emphasize that Fig.~\ref{fig:animated} is not more but a monitor to visualize what steps have led to the training result of the \gls{NN}.
This information can help to interpret both the features of the input space and the \gls{NN} sensitivity to it.
A different \gls{NN} configuration might reveal a different sensitivity to any of the $\langle t_{i}\rangle$.
Also there is no claim of proof that the increase in $\langle t_{x_1,x_2}\rangle$ causes the increase in the \gls{AUC}.

\section{Application to a benchmark task from high-energy particle physics}
\label{sec:application_on_a_scenario_from_particle_physics}

\begin{figure}
\centering
\includegraphics[width=0.85\linewidth]{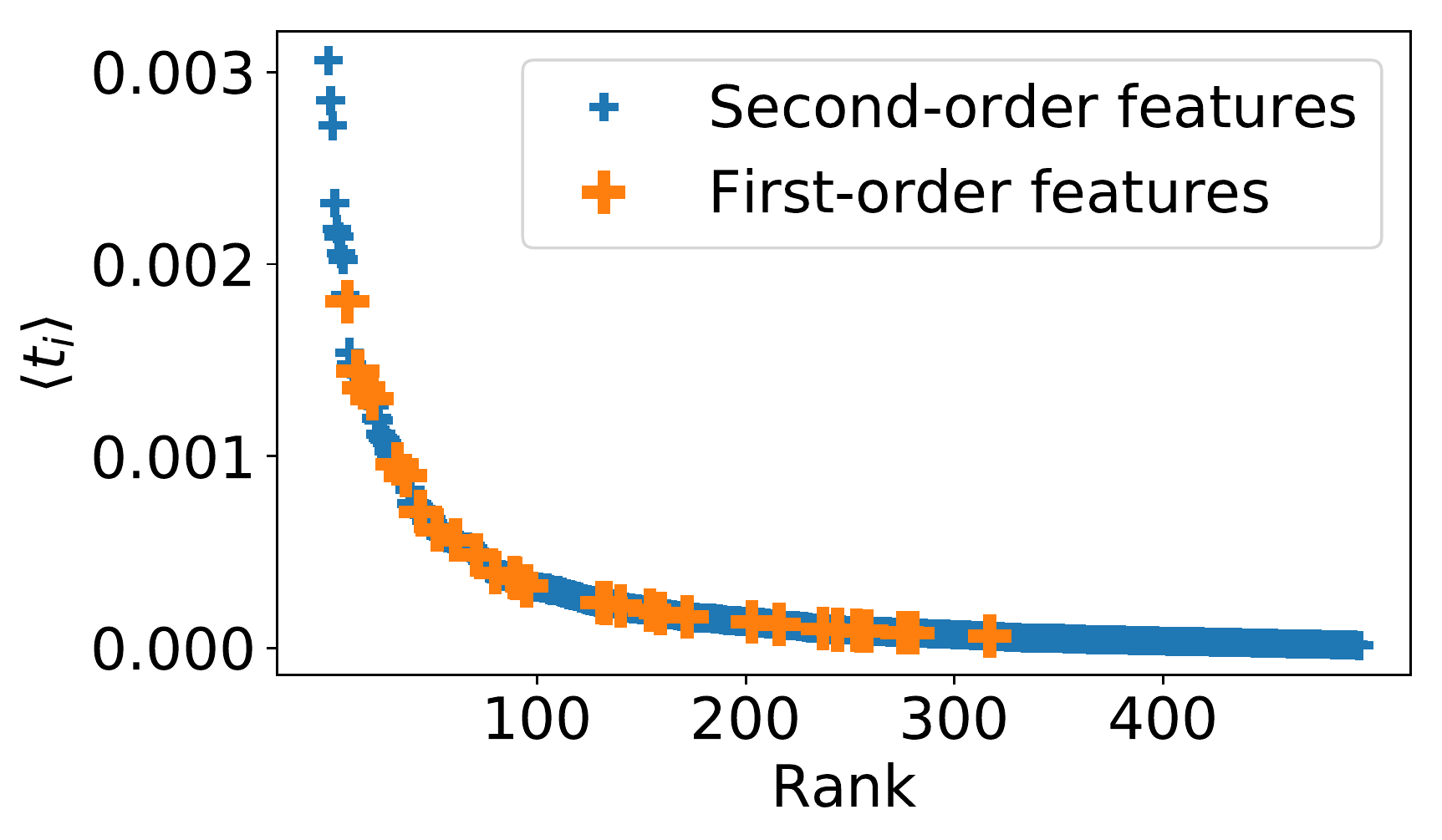}
\caption{Metrics $\langle t_i\rangle$, as defined in Eq.~\ref{eq:metric}, obtained from the $30$ inputs of the task discussed in section~\ref{sec:application_on_a_scenario_from_particle_physics}. The $\langle t_i\rangle$, have been ranked by value, in descending order. A color coding identifies (orange) first-order and (blue) second-order features.}
\label{fig:ranking_scores}
\end{figure}

\begin{figure}
\centering
\includegraphics[width=0.70\linewidth]{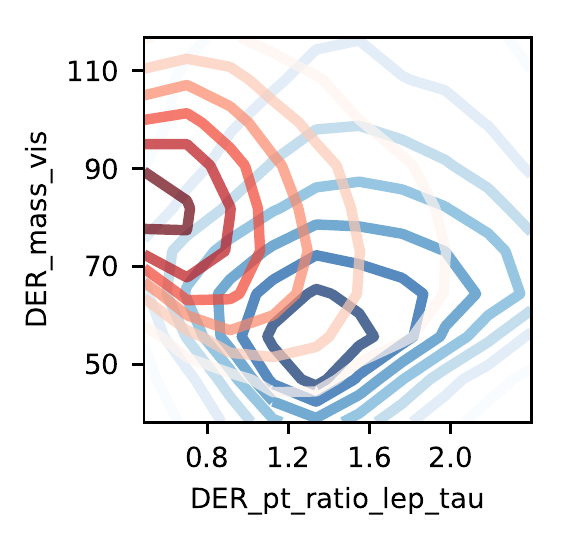}
\caption{Relation between the variables \texttt{DER\_mass\_vis} and \texttt{DER\_pt\_ratio\_lep\_tau}, as defined in~\cite{adam2014learning} and discussed in section~\ref{sec:application_on_a_scenario_from_particle_physics}, shown in a subset of the input space. The red (blue) contours correspond to the signal (background) class. Darker colors indicate a higher sample density. This relation is identified as the most influential feature after the \gls{NN} training.}
\label{fig:feature}
\end{figure}

\begin{figure}
\centering
\includegraphics[width=0.80\linewidth]{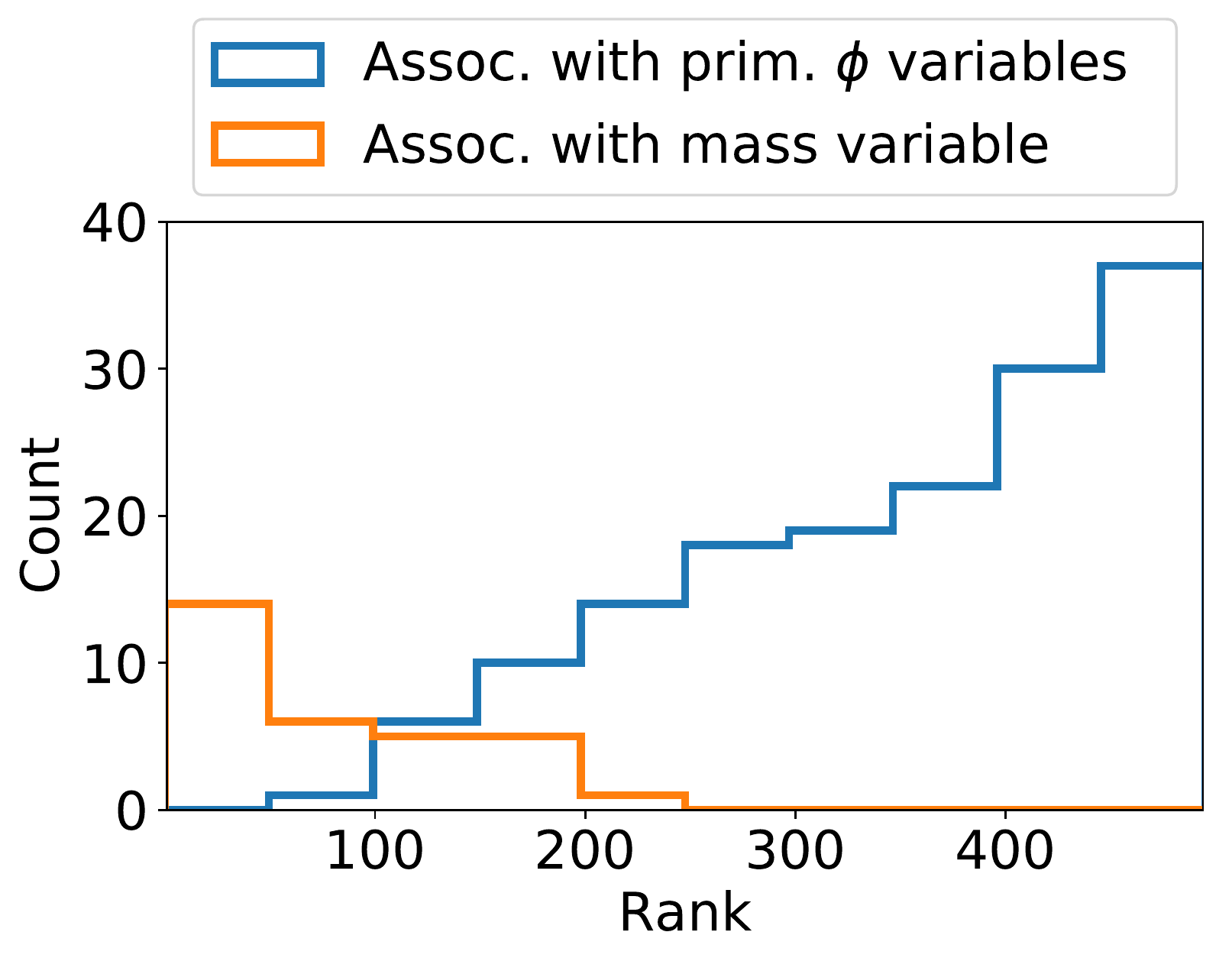}
\caption{Occurrence of features containing primitive $\phi$ variables and occurrence of \texttt{DER\_mass\_vis}, as discussed in section~\ref{sec:application_on_a_scenario_from_particle_physics}, in the ranked list of features.}
\label{fig:ranking_phi}
\end{figure}

In the following we are investigating the behavior of the $\langle t_i \rangle$ when applied to a more complex task, typical for data analyses in high-energy particle
physics.
For this purpose we are exploiting a dataset that was released in the context of the Higgs boson machine learning challenge~\cite{adam2014learning}, in 2014.
This challenge was inspired by the discovery of a Higgs particle in collisions of high-energy proton beams at the CERN LHC, in 2012~\cite{chatrchyan2012observation,aad2012observation}.
The search for Higgs bosons in the final state with two $\tau$ leptons~\cite{Chatrchyan2014nva,Aad2015vsa,Sirunyan2017khh} at the LHC has two main characteristics of relevance for this challenge:
\begin{itemize}
\item a Higgs boson will be produced in only a tiny fraction of the recorded collisions.
\item there is no unambiguous physical signature to
distinguish collisions containing Higgs bosons
(defining the signal class) from other collisions (defining the background
class).
\end{itemize}
Consequently, for such a search the signal needs to be inferred from a larger number of (potentially related) physical quantities of the recorded collisions, using statistical methods, which makes the task suited also for \gls{NN} applications.
For the challenge a typical set of proton-proton collisions was simulated, of which only a small subset contained Higgs bosons in the final state with two $\tau$ leptons.
Important physical quantities to distinguish the signal and background classes are the momenta of certain collision products in the plane, transverse to the incoming proton beams; the invariant mass of pairs of certain collision products; and their angular position relative to each other and to the beam axis.
In the context of the challenge the values of 30 such quantities were released, whose names and exact physical meaning are given in~\cite{adam2014learning}.
Seventeen of these variables are basic quantities, characterizing a collision from direct measurements; the rest, like all invariant mass quantities, are called derived variables and computed from the basic quantities.
These derived variables have a high power to distinguish the signal and background classes.
Other variables like the azimuthal angle $\phi$ of single collision products in the plane transverse to the incoming proton beams have no separating power between the signal and background classes, due to the symmetry of the posed problem.
The task is solved by the same \gls{NN} model and training approach as described in section~\ref{sec:toy_scenarios}. Applied to all $30$ input quantities this results in an AUC of $0.92$ and an approximate median significance, as defined in~\cite{adam2014learning}, of $2.61$.
In total, the $30$ input quantities result in $495$ first- and second-order features.
For further discussion we rank these features according to their extracted influence on the \gls{NN} output, based on the values of the corresponding $\langle t_i \rangle$, in decreasing order.
In Fig.~\ref{fig:ranking_scores} the $\langle t_i \rangle$ for all features are shown, split into (orange) first- and (blue) second-order features.
The distribution shows a rapidly falling trend, suggesting that only a small number of the investigated features significantly contributes to the solution of the task.
The most important input variable is identified as the invariant mass calculated from the kinematics of two distinguished particles in the collision, the identified hadronic $\tau$ lepton decay and the additional light flavor lepton, associated with a leptonic decay of the $\tau$ lepton, \texttt{DER\_mass\_vis}, as defined in~\cite{adam2014learning}.
This variable also belongs to the most important quantities to identify Higgs particles in the published analyses~\cite{Chatrchyan2014nva,Aad2015vsa,Sirunyan2017khh}, with a strong relation to the invariant mass of the new particle.
It is a peaking unimodal distribution in the signal class, with a broader distribution, peaking in a different position, in the background class.
Among the $10$ most influential features, it appears as the most influential first-order feature (in position $10$), reflecting the difference in the position of the peak in the signal and background classes, and as part of six further second-order features, including the auto-correlation (in position $6$), characterizing the difference in the width of the peak in the signal and background classes.
The \gls{NN} is thus able to identify the most important features of \texttt{DER\_mass\_vis}: its peak position and width.
The usage of this variable in a \gls{NN} analysis requires a good understanding not only of the marginal distribution but also of all relevant relations to other variables, which should be reflected in the uncertainty model.
The most influential feature is found to be the relation of \texttt{DER\_mass\_vis} with the ratio of the transverse momenta of the two particles that enter the calculation of this variable, named \texttt{DER\_pt\_ratio\_lep\_tau}.
This feature is shown in Fig.~\ref{fig:feature}, visualizing the gain of the relation over a pure marginal distribution on each individual axis.
Features related to $\phi$ on the other hand are consequently ranked to the end of the list, as can be seen from Fig.~\ref{fig:ranking_phi}, with the first occurrence in position $82$.
Apart from \texttt{DER\_mass\_vis} only eight more inputs, which are all well motivated from the physics expectation, contribute to the upper $\SI{5}{\percent}$ of the most influential features.
When exposed to only these nine input quantities the \gls{NN} solves the task with an \gls{AUC} and \gls{ROC} curve identical to the one that we observe, when using all 30 input quantities, within the numerical precision, indicating the potential to reduce the input space from $30$ to $9$ dimensions without significant loss of information.
We refrain from a more detailed analysis of the complete list of features, which quickly turns very abstract and cannot be fully appreciated without deeper knowledge of the exact physical meaning of the input quantities.
We conclude that the metric of Eq.~\ref{eq:metric} allows for a detailed understanding of the role of each input quantity - even without knowing their exact meaning - and quantitatively confirms the intuition of the high-energy particle physics analyses that have been performed during the search for the Higgs boson in 2012 and afterwards.
We would like to emphasize that the reduction of the dimension of the input space (in the demonstrated case from 30 to 9), which can be achieved also by other methods, like the principal component analysis~\cite{abdi2010principal}, is not the main goal of our investigation. The main goal is an improved and more intuitive understanding of the features of the input space and the sensitivity of the \gls{NN} output on it.

\section{Summary}
\label{sec:summary}

We have discussed the usage of the coefficients $t_i$ from a Taylor expansion in each element of the input space $\{x_j\}$ to identify the characteristics of the input space with the  largest influence on the \gls{NN} output.
For practical reasons we have restricted the discussion to the expansion up to second order, concentrating on the characteristics of marginal distributions of input elements, $x_j$, or relations between them, referred to as first- and second-order features.
We propose the arithmetic mean of the absolute value of a corresponding Taylor coefficient $\langle t_i \rangle$, built from the whole input space, as a metric to quantify the influence of the corresponding feature on the \gls{NN} output.
We have illustrated the relation between features and corresponding $\langle t_i \rangle$ with the help of simple tasks emphasizing single features or relations between them.
Evaluating the $\langle t_i \rangle$ at each step of the \gls{NN} training allows for the analysis and monitoring of the learning process of the \gls{NN}.
Finally we have applied the proposed metrics to a more complex task common to high-energy particle physics and found that the most important features, known from physics analyses are reliably identified, while features known to be irrelevant are also identified as such.
We consider this as the first step to identify those characteristics of the \gls{NN} input space that have the largest influence on the \gls{NN} output, in the context of tasks, typical for high-energy particle physics experiments.
As shown for the example in section~\ref{sec:application_on_a_scenario_from_particle_physics} these most influential characteristics may well correspond to relations between different inputs or auto-correlations, and not just to the marginal distribution of single inputs.
In subsequent steps the quantification of systematic uncertainties in the \gls{NN} inputs can be concentrated on those most relevant inputs.

\newpage
\bibliographystyle{splncs}
\bibliography{citations}

\begin{thebibliography}{10}

\bibitem{junk1999confidence}
Junk, T.:
\newblock Confidence level computation for combining searches with small
  statistics.
\newblock Nuclear Instruments and Methods in Physics Research \textbf{434}(2)
  (1999)  435

\bibitem{read2002presentation}
Read, A.L.:
\newblock {Presentation of search results: the CLs technique}.
\newblock Journal of Physics G: Nuclear and Particle Physics \textbf{28}(10)
  (2002)  2693

\bibitem{atlas2011procedure}
{The ATLAS and CMS collaborations}:
\newblock {Procedure for the LHC Higgs boson search combination in summer
  2011}.
\newblock Technical report, ATL-PHYS-PUB-2011-011, CMS NOTE 2011/005 (2011)

\bibitem{belforte2012combined}
{The CMS collaboration}:
\newblock {Combined results of searches for the Standard Model Higgs boson in
  $pp$ collisions at $\sqrt{s}$ = 7 TeV}.
\newblock Phys. Lett. B \textbf{710} (2012) ~26

\bibitem{cowan2011asymptotic}
Cowan, G., Cranmer, K., Gross, E., Vitells, O.:
\newblock Asymptotic formulae for likelihood-based tests of new physics.
\newblock Eur. Phys. J. C \textbf{71}(2) (2011)  1554

\bibitem{metodiev2017weakly}
Metodiev, E., Nachman, B., Thaler, J.:
\newblock {Classification without labels: Learning from mixed samples in high
  energy physics}.
\newblock arXiv:1708.02949 (2017)

\bibitem{dery2017weakly}
Dery, L.M., Nachman, B., Rubbo, F., Schwartzman, A.:
\newblock Weakly supervised classification in high energy physics.
\newblock Journal of High Energy Physics \textbf{2017}(5) (2017)  145

\bibitem{komiske2018weakly}
Komiske, P., Metodiev, E., Nachman, B., Schwartz, M.:
\newblock Learning to classify from impure samples with high-dimensional data.
\newblock arXiv:1801.10158 (2018)

\bibitem{cohen2018weakly}
Cohen, T., Freytsis, M., Ostdiek, B.:
\newblock {(Machine) Learning to Do More with Less}.
\newblock arXiv:1706.09451 (2018)

\bibitem{louppe2017learning}
Louppe, G., Kagan, M., Cranmer, K.:
\newblock Learning to pivot with adversarial networks.
\newblock In: Advances in Neural Information Processing Systems. (2017)  982

\bibitem{deoliveira2017flattening}
de~Oliveira, L., Kagan, M., Mackey, L., Nachman, B., Schwartzman, A.:
\newblock {Jet-Images -- Deep Learning Edition}.
\newblock arXiv:1511.05190 (2017)

\bibitem{chang2018flattening}
Chang, S., Cohen, T., Ostdiek, B.:
\newblock What is the machine learning?
\newblock arXiv:1709.10106 (2018)

\bibitem{bach2015pixel}
Bach, S., Binder, A., Montavon, G., Klauschen, F., M{\"u}ller, K.R., Samek, W.:
\newblock On pixel-wise explanations for non-linear classifier decisions by
  layer-wise relevance propagation.
\newblock PloS one \textbf{10}(7) (2015)

\bibitem{montavon2017explaining}
Montavon, G., Lapuschkin, S., Binder, A., Samek, W., M{\"u}ller, K.R.:
\newblock Explaining nonlinear classification decisions with deep taylor
  decomposition.
\newblock Pattern Recognition \textbf{65} (2017)  211

\bibitem{abadi2016tensorflow}
Abadi, M., Agarwal, A., Barham, P., Brevdo, E., Chen, Z., Citro, C., Corrado,
  G.S., Davis, A., Dean, J., Devin, M.,  et~al.:
\newblock {Tensorflow: Large-scale machine learning on heterogeneous
  distributed systems}.
\newblock arXiv preprint arXiv:1603.04467 (2016)

\bibitem{paszke2017automatic}
Paszke, A., Gross, S., Chintala, S., Chanan, G., Yang, E., DeVito, Z., Lin, Z.,
  Desmaison, A., Antiga, L., Lerer, A.:
\newblock {Automatic differentiation in PyTorch}.
\newblock (2017)

\bibitem{bergstra2010theano}
Bergstra, J., Breuleux, O., Bastien, F., Lamblin, P., Pascanu, R., Desjardins,
  G., Turian, J., Warde-Farley, D., Bengio, Y.:
\newblock {Theano: A CPU and GPU math compiler in Python}.
\newblock In: Proc. 9th Python in Science Conf. (2010) ~1

\bibitem{kingma2014adam}
Kingma, D., Ba, J.:
\newblock Adam: A method for stochastic optimization.
\newblock arXiv preprint arXiv:1412.6980 (2014)

\bibitem{chollet2015keras}
Chollet, F.,  et~al.:
\newblock Keras.
\newblock \url{https://keras.io} (2015)

\bibitem{adam2014learning}
Adam-Bourdarios, C., Cowan, G., Germain, C., Guyon, I., Kegl, B., Rousseau, D.:
\newblock {Learning to discover: the Higgs boson machine learning challenge}
  \url{https://higgsml.lal.in2p3.fr/documentation/} Visited on January 3, 2018.

\bibitem{chatrchyan2012observation}
{The CMS collaboration}:
\newblock {Observation of a new boson at a mass of 125 GeV with the CMS
  experiment at the LHC}.
\newblock Phys. Lett. B \textbf{716}(1) (2012) ~30

\bibitem{aad2012observation}
{The ATLAS collaboration}:
\newblock {Observation of a new particle in the search for the Standard Model
  Higgs boson with the ATLAS detector at the LHC}.
\newblock Phys. Lett. B \textbf{716}(1) (2012) ~1

\bibitem{Chatrchyan2014nva}
{The CMS collaboration}:
\newblock {{Evidence for the 125 GeV Higgs boson decaying to a pair of $\tau$
  leptons}}.
\newblock JHEP \textbf{05} (2014)  104

\bibitem{Aad2015vsa}
{The ATLAS collaboration}:
\newblock {{Evidence for the Higgs-boson Yukawa coupling to $\tau$ leptons with
  the ATLAS detector}}.
\newblock JHEP \textbf{04} (2015)  117

\bibitem{Sirunyan2017khh}
{The CMS collaboration}:
\newblock {{Observation of the Higgs boson decay to a pair of $\tau$ leptons
  with the CMS detector}}.
\newblock Phys. Lett. B \textbf{779} (2018)  283

\bibitem{abdi2010principal}
Abdi, H., Williams, L.J.:
\newblock Principal component analysis.
\newblock Wiley interdisciplinary reviews: computational statistics
  \textbf{2}(4) (2010)  433

\end{thebibliography}

\end{document}